\documentclass[aps,prd,
nofootinbib,
preprintnumbers,
twocolumn,
floatfix
]{revtex4-1}

\usepackage{epsfig}
\usepackage{amsmath}
\usepackage{amssymb}
\usepackage{graphicx}
\usepackage{xcolor}

\def\e{\mbox{e}}

\def\gtap{\raisebox{-.55ex}{\rlap{$\sim$}} \raisebox{.4ex}{$>$}}
\def\gsim{\mathrel{\gtap}}

\begin{document}

\title{Non-Primordial Solar Mass Black Holes}

\author{Chris Kouvaris} 
\email{kouvaris@cp3.sdu.dk}
\affiliation{CP3-Origins \& Danish Institute for Advanced Study DIAS,\\
University of Southern Denmark, Campusvej 55, DK-5230 Odense M, Denmark} 
\author{Peter Tinyakov}
\email{petr.tiniakov@ulb.ac.be}
\affiliation{Service de Physique Th\'{e}orique, Universit\'{e} Libre
  de Bruxelles (ULB),\\CP225 Boulevard du Triomphe, B-1050 Bruxelles,
  Belgium}
\author{Michel H.G. Tytgat}
\email{mtytgat@ulb.ac.be}
\affiliation{Service de Physique Th\'{e}orique, Universit\'{e} Libre
  de Bruxelles (ULB),\\CP225 Boulevard du Triomphe, B-1050 Bruxelles,
  Belgium}

\preprint{ULB-TH/18-05}
\preprint{CP3-Origins-2018-014 DNRF90}
\begin{abstract} 
We propose a mechanism that can convert a sizeable fraction of neutron stars  into black holes with mass $\sim 1M_\odot$, 
too light to be produced via standard stellar evolution. 
We show that asymmetric fermionic dark matter of mass $\sim$~TeV, with attractive self-interaction within the range that alleviates the problems of collisionless cold dark matter,  can accumulate in a neutron star and collapse, forming a seed black hole that converts the rest of the star to a solar mass black hole. We estimate the fraction of neutron stars that can become black holes without contradicting existing neutron star observations. Like neutron stars, such solar mass black holes could be in binary systems, which may be searched for by existing and forthcoming gravitational wave detectors. The (non-)observation of binary mergers of solar mass black holes may thus test  the specific nature of the dark matter. 
\end{abstract}

\maketitle

\section{Introduction}
\label{sec:introduction}

Stellar mass black holes (BH) are formed in the collapse of 
 stars with masses larger than $\sim 20M_{\odot}$. For such heavy stars, the gravity overcomes the impediment of the Fermi pressure of nucleons at the core, leading to the formation of relatively heavy BH. The stars with lower masses roughly between $9M_{\odot} \lesssim M \lesssim 20M_{\odot}$ evolve into neutron stars (NS) which have masses from $1.5M_\odot \lesssim M \lesssim 2.5M_\odot$. Even lighter stars end up as white dwarfs (WD) with masses up to $1.5M_\odot$ (the Chandrasekhar limit). As the gravitational collapse is halted by the Fermi pressure of the electron gas if the mass is below the Chandrasekhar limit and by the nucleon Fermi pressure  at masses below $2-2.5M_\odot$, there is no way stellar evolution can lead to formation of a BH with the mass below $\sim 2M_\odot$ \cite{Rhoades:1974fn}.

The first detection of gravitational waves (GW) from merging binary BH (BBH) by LIGO \cite{Abbott:2016blz} opens up a unique opportunity to study the population of stellar mass black holes. So far, only heavy BH with masses in excess of $\sim 5 M\odot$ have been observed, in rough agreement with the above picture. Clearly, it is of a fundamental importance to check for (non)existence of light  ($\lesssim 2M_\odot$) BH. If such light BH are found, some new mechanism must be assumed to explain their formation. A possibility presently discussed in the literature is that 
 stellar mass black holes could be of primordial origin \cite{Bird:2016dcv,Clesse:2016vqa,Sasaki:2016jop}, being created in the collapse of large 
inhomogeneities in the early Universe \cite{Hawking:1971ei,Carr:1974nx}.

In this paper we argue that there exists at least one alternative option: light $O(M_\odot)$ BH may be created from neutron stars by accumulation of dark matter (DM). If a sufficient amount of DM is accumulated inside a NS, it may collapse into a mini-BH that eventually ``eats up'' the rest of the star, thus resulting in a conversion of a sizable fraction of the NS into an  $O(M_\odot)$ BH. Thus, a discovery of such low mass BH would not necessarily imply its primordial origin, but may instead point towards a particular nature of DM.
On the contrary, the absence of solar mass BH among merger events could set new and strict limits on the parameter space of the corresponding DM models. 

For our scenario to work the DM must have interactions with baryons, be non-annihilating and self-interacting. The non-annihilation of DM is naturally achieved in asymmetric DM models \cite{Nussinov:1985xr,Petraki:2013wwa} where the relic abundance of DM is determined by an initial asymmetry between the population of DM and its antiparticles in the Early Universe. These models are also theoretically motivated because of the possibility to have a common mechanism that could explain both the DM relic abundance and the baryon asymmetry of the Universe. 
The self-interaction of DM is widely discussed as a  cure for some of the possible problems of collisionless cold DM \cite{Spergel:1999mh}.
Numerical simulations~\cite{Dave:2000ar,Rocha:2012jg,Zavala:2012us,Vogelsberger:2012ku} suggest that they can be mitigated with the DM self-interaction cross section $\sigma$ satisfying $0.1 \text{cm}^2/\text{g}<\sigma/m<10\text{cm}^2/\text{g}$, $m$ being the DM mass (see \cite{Tulin:2017ara} for a review). 

In general, DM can have an effect on properties of compact stars such as white dwarfs (WD) and  NS~\cite{Bertone:2007ae,Kouvaris:2007ay,Kouvaris:2010vv,Kouvaris:2010jy,Capela:2013yf,Bramante:2015cua,Baryakhtar:2017dbj,Ellis:2018bkr}. Asymmetric DM can have an even more dramatic effect: due to the absence of DM annihilations, the accretion of such DM into NS leads to its accumulation and may result in the collapse into a microscopic BH which eventually destroys the rest of the star~\cite{Goldman:1989nd,Kouvaris:2011fi,McDermott:2011jp,Kouvaris:2012dz,Kouvaris:2011gb,Guver:2012ba,Bramante:2013nma,Bell:2013xk}. From the mere existence of nearby old NS severe constraints have been imposed on both bosonic asymmetric DM \cite{Kouvaris:2011fi,McDermott:2011jp} and on fermionic asymmetric DM with attractive self-interactions \cite{Kouvaris:2011gb}. In this paper we explore further the observational consequences of the second type of models, namely the possibility to create solar mass BH out of NS. We  identify the astrophysical conditions as well as the DM parameter space under which such BH could be produced. We also discuss the prospects for detection of binary systems of solar mass BH by present and future gravitational wave experiments.

\section{Formation of a Light Black Hole from a Neutron Star}
\label{sec:form-light-black}
Schematically, our mechanism of conversion of a NS into a solar-mass BH is as follows. The DM captured by the NS thermalizes with nucleons and forms a cloud in the star center. As the number of DM particles increases beyond  a certain critical value $N_{\rm cr}$ the DM's own gravity and self-attraction start to dominate over the external potential of the star, at which point the DM cloud starts to collapse. After that, if the number of particles becomes (or already is) larger than another critical number $N_{\rm Ch}$ needed to overcome the DM Fermi pressure --- the analog of the Chandrasekhar limit --- the collapsing cloud forms a mini-BH inside the NS which then consumes the rest of the star, turning it into a solar-mass BH. Thus, the main condition for conversion of a NS into a BH is that the number of DM particles accumulated in the NS lifetime exceeds both $N_{\rm cr}$ and $N_{\rm Ch}$. One also has to check that the thermalization and collapse do not take too long, but this turns out not to give additional constraints for the parameters we consider below. We now go briefly over the BH formation stages, most of which have been previously considered in the literature.

\paragraph{Dark Matter Capture and Accumulation. }
\label{sec:dark-matter-capture}
To capture DM in a NS, DM-nucleon interactions are necessary. The number of DM particles $N_{\text{acc}}$ that are accumulated within time $t$ by a NS, taking into account the relativistic effects, is given by~\cite{Goldman:1989nd,Kouvaris:2007ay} 
\begin{equation}
N_{\text{acc}}=
\sqrt{6\pi}\, {\rho_{\rm dm}\over mv}\,{RR_g\over 1-R_g/R }\,  f\, t
\label{eq:capture_gen}
\end{equation}
where $m$ is the DM mass, $\rho_{\rm dm}$ and $v$ are the DM density and velocity dispersion at the NS location, $R$ and $R_g$ the NS radius and its Schwarzschild radius and $f$ is a cross-section-dependent efficiency factor, $f=\sigma/\sigma_{\rm crit}$, with $\sigma_{\rm crit}= 0.45 m_nR^2/M\simeq 1.3 \times 10^{-45}{\rm cm}^2$ being the critical cross section above which on average every particle passing through the NS is scattered and captured. By definition, at $\sigma\geq \sigma_{\rm crit}$ the efficiency is 100\% ($f=1$). It follows from Eq.(\ref{eq:capture_gen}) that in a typical galactic environment with $\rho_{\rm dm}=0.3{\rm GeV/cm}^3$ and $v=220{\rm km/s}$ the total amount of DM accumulated over Gyr is $N_{\text{acc}}\simeq 10^{39} ({\rm TeV/m})$ assuming full efficiency, which corresponds to the total DM mass of $10^{-15} M_\odot$.  

The cross section of the DM-nucleon interactions $\sigma$ is severely constrained by direct detection experiments. The current limits on spin-independent interactions set by several experiments ~\cite{Akerib:2016vxi,Aprile:2017iyp}
imply $\sigma_{\rm SI}~\lesssim10^{-45}{\rm cm}^2 (m/{\rm TeV})$
at $m\gsim 100$~GeV, close to $\sigma_{\rm crit}$ for $m\sim {\rm TeV}$.\footnote{Limits on spin-dependent DM-nucleon interactions correspond to cross-sections $\sigma_{\rm SD} \gg  \sigma_{\rm crit}$ and so do not lead to more capture of DM by NS than allowed by current SI constraints. }

\paragraph{Thermalization.}
\label{sec:formation-bh-inside}

Once gravitationally bound, a DM particle continues to pass through a star each half-period, and after some time starts to orbit the star center inside the star, eventually thermalizing with nucleons and concentrating within the thermal radius 
\begin{equation}
r_{\rm th} = \left({15T\over 8\pi G \rho_c m}\right)^{1/2} \simeq 
8\,{\rm cm} \left({{\rm TeV}\over m}\right)^{1/2},
\label{therm_rad}
\end{equation}
where we have used typical NS core density $\rho_c=10^{15}{\rm g/cm}^3$ and temperature $T=10^5$~K. One may check that the thermalization time scale is short (typically hundreds of years) for the parameter region of interest \cite{Kouvaris:2010jy}.

\paragraph{Self-attraction and collapse.}
In
absence of self-attraction, the number of DM particles needed for collapse, $N_{\rm Ch}=5\times 10^{48} ({\rm TeV}/m)^3$, is much larger than can be accumulated in typical conditions unless the DM is very heavy,  $m\gsim 1000$~TeV. A more natural possibility is to assume that the DM possesses a Yukawa-type attractive self-interaction $V(r) = \alpha \exp (-\mu r)/r$, $\mu$ being the mediator mass.  In what follows we  consider 4 benchmark parameter values (see Table~\ref{table1} below), all lying in the range that alleviate the problems of collisionless DM. 

First we consider how the presence of the self-attraction modifies $N_{\rm cr}$ and $N_{\rm Ch}$. 
We consider now how the presence of the self-attraction modifies $N_{\rm cr}$ and $N_{\rm Ch}$. The critical number of particles $N_{\rm cr}$ above which the collapse begins can be estimated from the virial theorem assuming that the  DM is in equilibrium at the NS temperature $T$ and solving for the size $r$ of the cloud which we assume to be a uniform sphere for simplicity. The potential energy of DM cloud includes the external potential of the NS, self-gravitation and self-attraction terms. The virial theorem for the DM cloud reads
\[
2\langle E_k\rangle = \frac{8\pi}{5}  G\rho_c m r^2+ \frac{3GNm^2}{5r}
\]
\begin{equation}
+  {3N\alpha\e^{-\mu r_0 } \over 2\mu^2 r^3} \left( 3+3\mu r_0 + \mu^2 r_0^2\right)  ,
\label{virial}
\end{equation}
where $\langle E_k\rangle$ is the mean kinetic energy per particle, $r_0=n_0^{-1/3}= r(4\pi/3N)^{1/3}$ is the mean inter-particle distance, $n_0$ being the particle density. The last term requires an explanation. The  total contribution of the Yukawa interaction potential into the virial theorem reads $1/2 \langle \sum_{ik} (1+\mu r_{ik}) V (r_{ik})\rangle $ where $r_{ik}$ is the distance between particles $i$ and $k$. To get the contribution per particle (the last term in Eq.(\ref{virial})) we fix the particle $i$ somewhere in the cloud and replace the sum over $k$ by the integral over $r$ from $r_0$ to infinity, assuming the cloud size to be much larger than the Yukawa range $1/\mu$.  

In thermal equilibrium one has $2\langle E_k\rangle =3T$. With only the first term in Eq.(\ref{virial}) present, the r.h.s. changes from 0 to infinity and the solution for $r$ always exists --- this is the thermal radius (\ref{therm_rad}). When the self-gravitation and Yukawa terms are added with  coefficients that are small at small $N$, another (unstable) solution appears at small $r$ where these terms are singular. As $N$ grows the two solutions merge together and disappear, which signals the onset of collapse. We find this critical value $N_{\rm cr}$ numerically.

\paragraph{Overcoming Fermi pressure. }
As the collapse starts the DM cloud shrinks further, the energy of the cloud being evacuated through DM interactions with nucleons. Note that due to the virial theorem, the reduction in size heats up the DM particles which facilitates further energy transfer from DM to nucleons. The collapse time scale has been estimated in~\cite{Kouvaris:2012dz} and does not exceed a few thousand years in the worst case scenario for our parameters of interest. 

The collapse  may still be halted by the Fermi pressure if the number of DM particles is smaller than $N_{\rm Ch}$. In general, the Fermi-supported equilibrium configuration and the parameters at which it ceases to exist (i.e. the collapse to BH occurs) are determined by the solution to the Tolman-Oppenheimer-Volkoff equation. In the context of self-interacting DM  this problem was addressed in Ref.~\cite{Kouvaris:2015rea}. 

Qualitatively, the role of the self-interaction may be understood by considering the total energy of the cloud $E(r) = E_{\rm k} + E_{\rm pot}$ and looking for its local minimum as a function of $r$. In potential energy $E_{\rm pot}$ the contribution of the gravitational field of NS can in our case be neglected, while in the Yukawa term one may assume $\mu r_0\ll 1$. Moreover, the kinetic energy can be taken in the non-relativistic form $E_k = N^{5/3}/(mR^2)$. The equation $dE/dr=0$ has two solutions for $r$, a minimum and a maximum, which merge and disappear at a critical value of $N=N_{\rm Ch}$. The latter has been calculated in Ref.~\cite{Kouvaris:2015rea} and reads
\begin{equation}
N_{\rm Ch} = 0.3 \left({\mu\over m\sqrt{\alpha}}\right)^3 \left({M_{\rm Pl} \over m}\right)^3,
\label{eq:NCh}
\end{equation}
This is parametrically smaller than in the absence of the Yukawa attraction by the factor $(\mu/ m\sqrt{\alpha})^3$ which is small for our choice of parameters (note that the same combination controls the non-relativistic approximation). 
 
\paragraph{Conversion of NS into a BH. }
Any DM cloud inside a NS with  a number of DM particles larger than both $N_{\rm cr}$ and $N_{\rm Ch}$ will eventually collapse into a BH. As has been argued in~\cite{Kouvaris:2011fi}, if the  mass of the resulting BH  is smaller than $\sim 10^{-20} M_{\odot}$ it will evaporate due to the Hawking radiation faster than grow by accretion of NS matter, producing no observable effect. On the contrary, BH heavier than  $\sim 10^{-20} M_{\odot}$, which is the case for our choice of parameters, will grow very fast due to accretion, eventually destroying the star. More subtle issues regarding the effect of NS rotation and radiation from infalling matter on the growth of the BH have been addressed in~\cite{Kouvaris:2013kra}. In fact, due to rotation, a fraction of the NS mass might escape falling into the BH, so that the final BH mass might be somewhat smaller than the original mass of the NS. 
 
To conclude this section, we summarize the resulting numbers for our 4 benchmark cases in Table~\ref{table1}. All the criteria for the NS conversion into a BH  are satisfied for these parameters as soon as $N>N_{\rm Ch},\, N_{\rm cr}$. 
\begin{table}
\begin{tabular}{c||c|r|r|r|r|l}
\# & $\alpha$ & $\mu$ & $m$ &  $N_{\rm cr}$ & $N_{\rm Ch}$  & $M_{\rm Ch}$  \\
\hline
1  &  $10^{-4}$ & 1 MeV & 1 TeV & $3\cdot 10^{33}$& $6\cdot 10^{35}$ & $5\cdot 10^{-19} M_\odot$ \\
2  &  $10^{-3}$ & 10 MeV & 1 TeV & $5\cdot 10^{35}$ & $ 2\cdot 10^{37}$& $2\cdot 10^{-17} M_\odot$ \\
3  &  $10^{-3}$ & 1 MeV & 200 GeV & $1.3\cdot 10^{34}$ & $3\cdot 10^{38}$ & $5\cdot 10^{-17} M_\odot$ \\
4  & $10^{-4}$ & 1 MeV & 200 GeV &$3.7\cdot 10^{34}$ & $8\cdot 10^{39}$ & $2\cdot 10^{-15} M_\odot$ \\  
\end{tabular}
\caption{Benchmark values of Yukawa self-attraction parameters,  corresponding critical numbers $N_{\rm cr}$ and $N_{\rm Ch}$, and resulting mass of the mini-BH. 
\label{table1}
}
\end{table}

\section{Fraction of Collapsing Neutron Stars}
\label{sec:fract-coll-neutr}
The fate of a NS is determined by the total number of DM particles it has accumulated in its lifetime, which  according to Eq.~(\ref{eq:capture_gen}) is controlled by the DM density, velocity and cross section (through the efficiency $f$). Note that these parameters are degenerate, entering  in a single combination $\rho_{\rm dm}\sigma/v$. Since NS {\em are} observed in our galaxy, these parameters must be such that the observed NS survive. 
There are two potential areas of trouble: very old NS close to us and pulsars close to the galactic center where DM density is higher. In particular, J2124-3358 lies 270~pc from the Earth, with an age of 7.2~Gyr~\cite{Mignani:2003nw} and a core temperature of $2.2\times 10^6$K~\cite{Rangelov:2016syg}. There are also pulsars close to the galactic center: J1745-2900 is located at 0.1~pc from the center with an age of $9\times 10^3$~yr and a surface temperature of $\sim$ 1keV~\cite{Mori:2013yda}, corresponding to core temperature $\sim 10^{9}$~K. Other pulsars have been observed within 45~pc from the galactic
center. In particular, J1746-2856 is 1.2~Myr old~\cite{Johnston:2006fx}; the typical core temperature for this age is $\sim 10^7$~K. 

We need now to estimate the fraction of NS that collapse for a given set of parameters. For that,
we assume that the DM density in our galaxy follows a Burkert profile 
$\rho_{\rm Bur}=\rho_s (1+r/r_s)^{-1}\left [1+ (r/r_s)^2\right ]^{-1}$, with $\rho_s=3.15\text{GeV}/\text{cm}^3$ and $r_s=5$ kpc~\cite{Burkert:1995yz}. 
For the velocity distribution we assume a linear growth within 0.5\,kpc and a constant 220\,km/s at larger distances, which for numerical purpose we smooth with an hyperbolic tangent function. We also need the NS distribution within the Galaxy, which we assume to follow the stellar distribution. We adopt the bulge and a double disk distribution of Ref.\cite{McMillan:2011wd} with the best-fit parameters.

\begin{figure}[th]
\includegraphics[width=0.42
\textwidth]{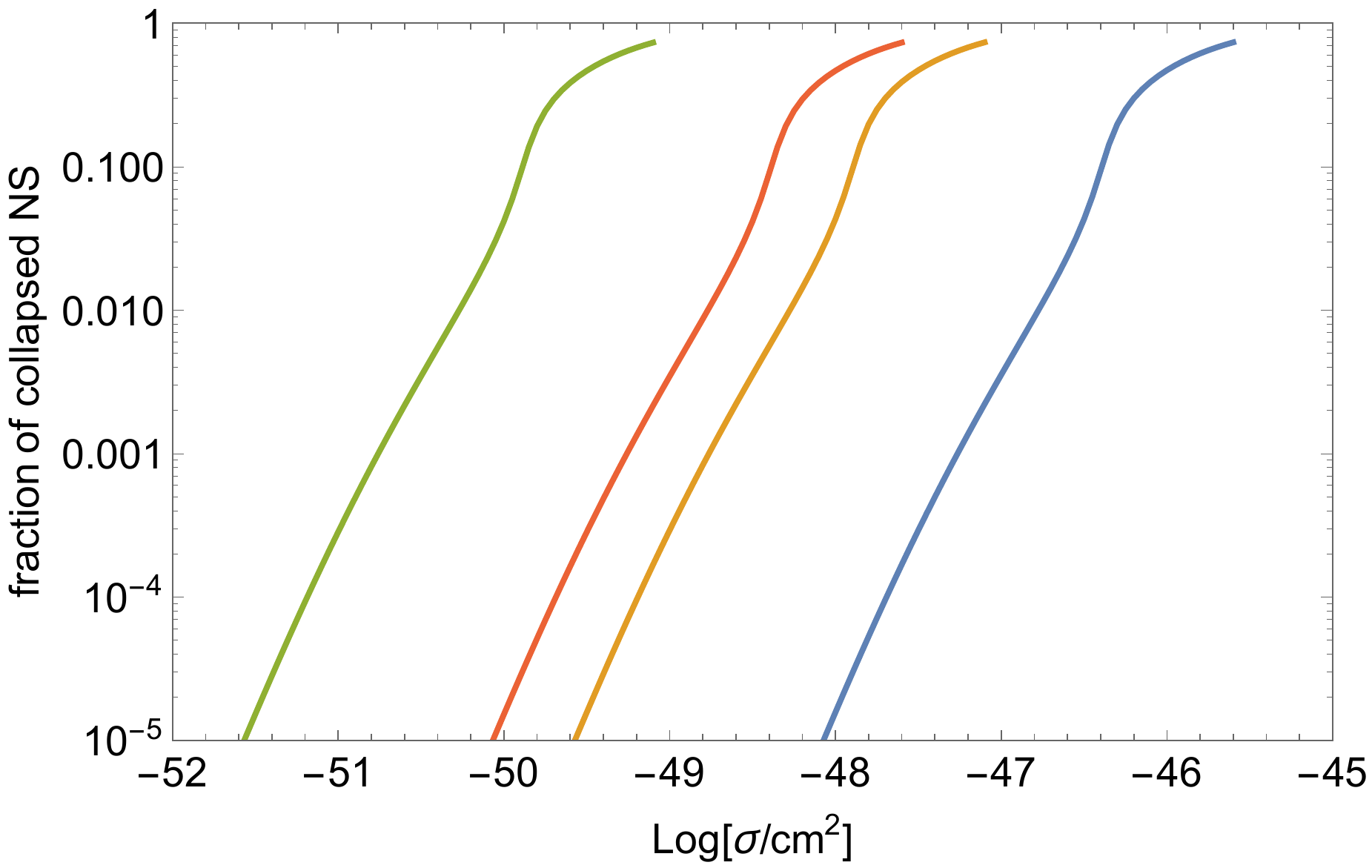}
\caption{Fraction of NS collapsed to BH in our Galaxy as a function of the DM-nucleon cross section for benchmark models 1--4 from left to right. 
\label{fig:fraction}
}
\end{figure}
When one increases the DM-nucleon cross section $\sigma$ (assuming $\sigma< \sigma_{\rm cr}$) with other parameters fixed, the fraction of NS in the Milky Way converted into BH increases as locations with lower $\rho_{\rm dm}/v$ begin to satisfy the collapse condition. Fig.~\ref{fig:fraction} shows this fraction as a function of $\sigma$ for 4  benchmark cases of Table~\ref{table1}. Here we have taken as average NS parameters the age of 5~Gyr and temperature of $10^5$~K. The curves stop as soon as one of the observed NS gets converted into a BH. They have the same shape as a result of the parameter degeneracy alluded to above. The corresponding maximum fraction of converted NS is $\sim 80\%$, based on the mere existence of J2124-3358. The exact numerical value of this maximum fraction should be taken with care as many factors have not been taken into account, like the NS distribution in core temperatures and ages and  uncertainties in the DM distribution. An extra uncertainty will arise when translating the results for  our galaxy into average numbers for a large volume. In any case, our estimate indicates that a sizeable fraction ($> 10\%$) can be achieved by this mechanism.

\section{Observational signatures}
\label{sec:observ-sign}

Consider the  gravitational wave signal produced by BH binary mergers in the mass range corresponding to neutron stars. At the time of this writing, one NS-NS merger has been observed by GW emission (GW170817)  \cite{TheLIGOScientific:2017qsa}. The total mass of this binary neutron star (BNS) system is $2.74^{+0.04}_{-0.01} M_\odot$ and its distance is $40^{+8}_{-14}$ Mpc. The rate of BNS mergers inferred is $1.5^{+3.2}_{-1.2}$ Myr$^{-1}$ Mpc$^{-3}$ ($90 \%$ C.L.). This rate is not inconsistent with early estimates \cite{Kalogera:2003tn}, but the uncertainties on the actual rate are quite large (see e.g. \cite{Belczynski:2017mqx}). To be concrete we will refer here to the rates quoted in the regularly updated review \cite{Aasi:2013wya}. At the horizon 2024, an advanced network of GW detectors is expected to observe about one BNS event per week. Specific numbers are shown in table \ref{table:GWdata} for LIGO/Virgo, as well as KAGRA (Kamioka Gravitational Radiation Antenna)
and the more future Einstein Telescope (ET). Obviously, if a fraction $>10\%$ of NS are converted into light BH, their merging will certainly be easily observed.  The non-observation of anomalous BH-BH mergers over a period of time $\Delta t$ would bound  their rate  to $R \leq 2.3/\Delta t$ at $90\%$C.L., or  their fraction to $\leq 2.3/N_{\rm BNS}$  where  $N_{\rm BNS}$ is the observed number of BNS events. Ten years of observations with an advanced network could constrain a fraction potentially as low as $ \lesssim 10^{-2}-10^{-3}$ (respectively  $\lesssim 10^{-4}-10^{-8}$ for the ET) excluding corresponding DM candidates, see Fig.~\ref{fig:fraction}. 
\begin{center}
\begin{table}[ht]
\begin{tabular}{c | c | c}
Detectors & BNS range (Mpc) & BNS detections (per year)  \\
\hline
LIGO/Virgo & $105/80$ &  $4-80 \;(2020+)$\\
KAGRA & $100$ & $11-180\; (2024+)$\\
ET  & $\sim 5\cdot 10^3\; (z\approx 2)$	& ${\cal O}(10^3-10^7)$\\
\hline
\end{tabular}
\caption{GW Detectors prospects for BNS detection. The sensitivities are expressed in terms of the BNS distance range, assuming a $1.4 M_\odot + 1.4 M_\odot$ binary system \cite{Aasi:2013wya}. 
For the Einstein Telescope (ET) the numbers are based on a rate of $0.1-6$ BNS events/Myr/Mpc$^{3}$ \cite{ET:2011}.}
\label{table:GWdata}
\end{table}
\end{center}

These figures are only tentative, but they reveal that there is a potential for testing such a scenario in the near future. One may still question whether it will be possible to differentiate a BNS merger from that of a BBH. There are several aspects to this question, most of which go beyond the scope of this work. For instance, a BNS merger detection may be complemented by other observations, in particular in photons (from radio wavelengths to gamma-rays), so observing a merger with mass $\sim 1.5 M_\odot + 1.5 M_\odot$ and no associated gamma-rays could be a signature of an exotic event. Another aspect is that in our scenario NS may transform into BH only provided they are old enough, while BNS mergers are expected to trace the star formation history, so redshift distributions of events in the two cases should be different. 
Perhaps a more immediate and simpler issue is whether the GW signatures of a BNS and of a BBH of same mass differ at all in the detectors. To estimate the strength of the signal from BBH merger and to compare it with the one from BNS systems for given detector sensitivities, one may use the so-called characteristic strain $h_c$, which is defined as \cite{Moore:2014lga} 
$h_c(f) = \sqrt{2 f^2 /\dot f}\, h_0(f)$
where $f$, $\dot f$ are the frequency of the GW and its time derivative and $h_0$ is the root mean square strain. This quantity is designed to capture both an estimate of the instantaneous amplitude and of the duration of the signal, as both are relevant in determining the signal-to-noise ratio (SNR) of a possible event. 

\begin{figure}[t]
\includegraphics[width=0.85
 \columnwidth]{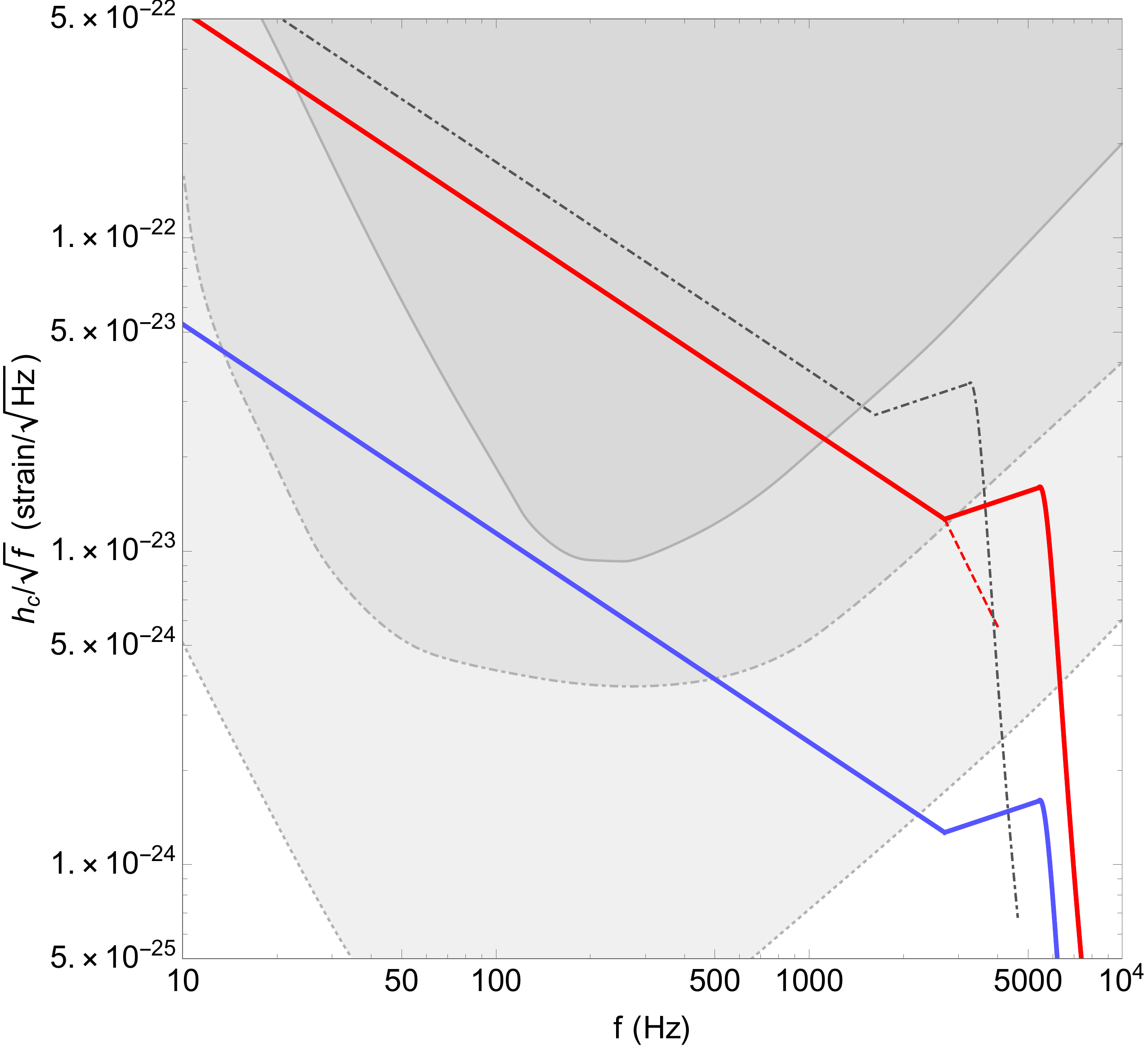}
\caption{\label{fig:spectra} Spectrum of GW from a $(1.5 + 1.5) M_\odot$ BBH at $40$ Mpc (red solid).  The spectrum of a corresponding BNS is schematically depicted by the break (red dashed). Also shown are a $(1.5 + 1.5) M_\odot$ BBH at $400$ Mpc (blue solid) and  a $(2 + 2) M_\odot$ BBH at $40$ Mpc (grey dot-dashed). The sensitivity curves are for to LIGO2017 (black solid), LIGO design (black dot-dashed) and ET design (black dotted).}
\end{figure}

The mergers of NS-NS binary and BH-BH binary of a similar mass 
differ most significantly by two features. First, the maximum typical frequency of peak signal is expected to be related to the characteristic size $R$ of the system at the moment of merging, with 
$\omega \approx M/R^3$. For identical mass, a black hole would be more compact than the corresponding NS by a factor of $\sim 3$, so the merging of  a BNS system should, a priori, occur at a lower frequency than its BH counterpart. Second, a BBH merger is observed to release most of its energy during the merger, with a corresponding peak in its spectral signature.
The spectral signature of a BNS merger is more complex.  Depending on the properties of the system (e.g. whether or not a black hole is formed in the process, etc.) it may be altogether absent (see e.g. \cite{Kiuchi:2010ze,Clark:2015zxa}). Estimates based on numerical simulations support these expectations. To be concrete, using the template for BBH mergers of Ref.\cite{Ajith:2007kx}, we get that a system with $(1.5+1.5) M_\odot$ would peak around $6$ kHz, while, for instance Ref.\cite{Clark:2015zxa} predicts a signal that would peak at frequencies between $2$ and $3$ kHz. Other studies lead to a broader range, but also less significant peak \cite{Kiuchi:2010ze}. For the sake of illustration, we show in Fig.\ref{fig:spectra} the spectral signatures of  possible BBH candidates with mass in the NS range, together with the sensitivity of current and future detectors. 
Clearly, the distinction between similar mass BNS and BBH mergers should be possible if the final stage of the merger falls within the detector sensitivity. 

\section{Conclusion}
\label{sec:conclusion}
BH created from the collapse of a star are deemed to be heavier than $\sim 2M_\odot$. We have introduced a mechanism through which a sizeable fraction of the NS can be converted into lighter $\sim 1M_\odot$ BH, which shows that, if such light BH are ever observed, they do not have to be necessarily of primordial origin. Our scenario involves  DM in the form of fermionic particles with masses around or below TeV and cross sections with nucleons characteristics of weakly interacting massive particles. Crucially, the DM particle must also be asymmetric and have attractive self-interactions, all properties that are theoretically well motivated and much discussed in the literature, either in relation with the baryon asymmetry of the Universe or to address possible small scale issues of collisionless cold dark matter. 
Primordial BH constitute a plausible alternative to DM particles. It is thus ironic that DM particles may lead to objects similar to primordial BH, albeit within a specific mass range. 

As a proof by example, we have considered 4 specific benchmark DM particle candidates. These do not exclude the possibility that a similar scenario
may be constructed for other DM candidates, fermionic or bosonic, provided that sufficient amount of DM can be accumulated by a NS and then made to collapse into a BH. 
We have also put forward the possibility of detecting light BBH systems using existing and forthcoming GW detectors and briefly discussed the issue of disentangling the signal from BNS and BBH systems of similar mass. 

Our work may and should be extended in several directions.  The precise range of BH masses that can be created from collapse of NS remains to be established. Indeed, it is conceivable that our mechanism leads to BH that are even lighter than a solar mass, for instance due  to the rapid rotation or to the magnetic field of the progenitor NS. 
Also, our estimate for the fraction of NS that may be converted into BH depends on both the DM properties and on the NS environment. We have considered a few benchmark DM candidates and have made specific assumptions regarding the distribution of both NS and DM in galaxies. Determining more precisely both the fraction of NS that can be converted into BH and the ability of GW detectors to disentangle BBH from their BNS counterparts would allow to set new and stringent constraints on the properties of dark matter.

\acknowledgments

The work of P.T.  is supported by the IISN; the work of M.T. by the IISN and the FNRS. CK is partially funded by the Danish National Research Foundation, grant number DNRF90, and by the Danish Council for Independent Research, grant number DFF 4181-00055. P.T. and M.T. would like to thank Raghuveer Garani for  useful discussions.

\end{document}